\def\leftdisplay$${\leftline{\noindent\ifleqno\eqn
  \indent\fi$\displaystyle{\eq}\ifeqno\hfill\ifleqno\else\eqn\fi\fi$}$$}
\newif\ifeqno \newif\ifleqno \everydisplay{\displaysetup}
\def\displaysetup#1$${\displaytest#1\eqno\eqno\displaytest}
\def\displaytest#1\eqno#2\eqno#3\displaytest{%
  \if!#3!\ldisplaytest#1\leqno\leqno\ldisplaytest
  \else\eqnotrue\leqnofalse\def\eqn{#2}\def\eq{#1}\fi\leftdisplay$$}
\def\ldisplaytest#1\leqno#2\leqno#3\ldisplaytest{\def\eq{#1}%
  \if!#3!\eqnofalse\else\eqnotrue\leqnotrue\def\eqn{#2}\fi}
\def\ref{\par\noindent\hangindent=0.7 true cm
         \hangafter=1}
\magnification=\magstep0
\hsize=16.0 cm
\vsize=24.5 cm
\baselineskip=12 pt plus 1 pt minus 1 pt 
\parindent=0.0 cm
\hoffset=0.0 cm
\voffset=-0.0 cm
\parskip = 3 pt
\def\centreline{\centerline}
\font\twelvebf=cmbx10 at 12truept
\input epsf
\centreline{\twelvebf Response of a Spaceborn Gravitational Wave Antenna to Solar Oscillations}
\vskip 20pt
\centreline{I.W.Roxburgh$^{1,2}$, A.G.Polnarev$^1$, G.Giampieri$^{1,3}$, S.V.
Vorontsov$^{1,4}$}
\vskip 8pt
\centreline{$^1$Astronomy Unit, Queen Mary, University of 
London, London E1 4NS, UK}
\centreline{$^2$DESPA, Observatoire de Paris, 92155 Meudon, France}
\centreline{$^3$Department of Physics, Imperial College, University of London, London SW7 2AZ}
\centreline{$^4$Institute of Physics of the Earth, Moscow, 123810, Russia} 
\vskip20pt
\centreline{\bf Abstract}

We investigate the possibility of observing very small amplitude low 
frequency solar oscillations with the proposed laser interferometer space 
antenna LISA.  For frequencies 
below $\sim 2\times 10^{-4}$ Hz the dominant contribution is from 
the near zone time 
dependent gravitational quadrupole moments associated with the normal 
modes of 
oscillation.  For frequencies $\nu$ above $\sim 3\times 10^{-4}$ Hz 
the dominant contribution is from gravitational radiation generated by 
the quadrupole
oscillations which is larger than the Newtonian signal by a factor 
$\sim (2 \pi r \nu/ c)^4$, where $r$ is the distance to the Sun, 
and $c$ is the velocity of light.

The low order solar quadrupole pressure and gravity oscillation modes have not 
yet been detected above the solar background by helioseismic velocity and 
intensity measurements.  Our estimates of the amplitudes needed to give a 
detectable signal on a LISA type space laser interferometer imply
surface velocity amplitudes on the sun of the order of $1-10$ mm/sec in the frequency range $1 - 5~10^{-4}$Hz. Such surface velocities are below the current sensitivity limits on helioseismic measurements. If modes exist with frequencies and amplitudes in this range they could be detected with a LISA type laser interferometer.
\vskip 15pt
{\bf 1. Introduction}

The proposed ESA/NASA proposed gravitational wave space interferometer 
LISA (ESA 1998) consists  
of three spacecraft at the vertices of an equilateral triangle of
sides $5\times10^9$ metres;
the system is maintained in this configuration by arranging that the plane of the detectors has an inclination of $60^o$ to the ecliptic and counter rotates with the same period as it orbits the Sun. Any two arms constitute a Michelson type interferometer, a mother spacecraft sends a laser beam to the other two 
satellites where the signal is 
coherently transponded back: the interferometer readout is 
obtained by interfering the incoming signal with the outgoing one 
and comparing the fractional change in phase shift between the two arms.
These interferometers will be sensitive to the passage of 
gravitational waves in the frequency range between $10^{-4}$ and 1 Hz, 
a range currently inaccessible on the ground due to seismic noise. 
The primary goal of these experiments is to detect waves from individual sources
(close binary systems, neutron star or black hole coalescence) and any 
stochastic background due to the superposition of 
waves emitted by binary systems, - and possibly from the early Universe. 

Current studies of the sensitivity of the LISA experiment (ESA 1998) indicate that the instrumental noise, due dominantly to the residual uncompensated accelerations, 
$\sim 3\times 10^{-18} /\sqrt{\rm Hz}$ at $10^{-4}$ Hz. In addition there is likely to be background "confusion noise" from binary systems which has a comparable magnitude at $10^{-4}$ Hz. (Bender and Hils 1997). When due 
allowance is made for the 
angle between the arms of the detector and the unknown direction of a source of gravitational radiation, the dimensionless strain that could be detected at a signal to noise of 5 is estimated to be $\sim 7\times10^{-21}$ with 1 year's observations.

A LISA type interferometer is sensitive to any variations in the 
gravitational field in the frequency range $10^{-4} - 1$Hz.  Since the Sun is known to be oscillating in normal modes of small amplitude with frequencies  in this range, the oscillating external gravitational field would be detectable by LISA if the amplitudes are large enough (cf Cutler and Lindblom 1996, Polnarev et al 1996, Roxburgh et al 1999). Only quadrupole modes are likely to be large enough to be detected at 1 a.u. and are the modes studied here. In addition to the time dependent external Newtonian gravitational field the 
time varying gravitational quadrupole moments also generate gravitational waves, which could give a detectable signal on a LISA type interferometer. At a distance of $r=1$ a.u. $2\pi\nu r/c > 1$ for $\nu > \nu_r \sim 3\times10^{-4}$ Hz, so we anticipate that gravitational radiation will be important at  frequencies greater than this. 

The first evidence of surface layer solar oscillations dates back to the work of Leighton et al (1962). Low order gobal oscillations, which are the oscillations of interest here, were detected by Claverie et al (1979)
 and Grec et al (1980). as resolved peaks in the power spectrum of a time series of measurements of the Doppler shift of a K and Na line using the integrated light from the Sun. As a result of ground based and space based 
observational programmes upwards of $10^7$ oscillation modes have been identified in the frequency range $10^{-3} - 10^{-2}$ Hz.; knowledge of these frequencies has been used to infer the acoustic and dynamical structure of the Sun (pressure, density and rotation as a function of radius), placing 
constraints on  the physics of the solar interior and on models of solar 
(and thereby stellar) evolution. 

The (p-modes) that have been detected are concentrated in the solar surface layers so the amplitude of the oscillating gravitational field is very small.  Modes of lower frequency $< 5\times 10^{-4}$ Hz have amplitudes that are relatively much larger in the core and correspondingly a larger external gravitational field for a given surface amplitude; since they are more sensitive to the core, knowledge of their frequencies would place much tighter constraints on the structure of the very central solar core. Such modes (g-modes and low order p-modes) have not yet been detected above the solar background noise which increases at low frequencies and great effort is currently being expended in the search for such modes. 

Here we investigate the possibility of detecting such low frequency quadrupole oscillations with a LISA type laser interferometer, including both the Newtonian near zone perturbations and the associated gravitational wave emission in the frequency range $3\times 10^{-5} - 10^{-3}$ Hz. We compare the gravitational signals detectable by laser interferometry with the velocity signals detectable by whole disc helioseismolgy, and demonstrate that
low frequency quadrupole oscillations with surface velocity amplitudes below 
current helioseimic limits could nevertheless be large enough to be detectable 
by LISA.  If such modes are first detected by helioseismic techniques then the measured frequencies (and predicted power) would provide a valuable calibration tool for LISA; if they have not been detected by helioseismic means then one can look upon LISA as a potential telescope for studying the deep solar interior.

The plan of this paper is as follows: first we discuss the properties of the solar oscillations and the quantify the relationship between the surface radial amplitude of an oscillation, its quadrupole moment and its horizontal amplitude.  We then express the external gravitational field in terms of the quadrupole moment tensor expressed as a sum over a set of basis tensors (corresponding to surface harmonics) which enables us to relate the generation of gravitational waves to the surface amplitude of the oscillation. 
We then determine the response of a LISA type detector to both the time dependent Newtonian field and to the associated gravitational waves for a given oscillating quadrupole moment; above $\sim 3\times 10^{-4}$ Hz the signal is dominated by gravitational waves.  Next we determine the magnitude of the velocity signal from each mode in terms of surface amplitude. We then consider the background noise for both velocity and gravitational detectors and hence the
signal to noise in both gravity and velocity for a given assumed frequency resolution.  The ratio of these S/N is then independent of the assumed surface amplitude of the oscillation and if greater than 1 the modes are easier to detect by a gravitational laser interferometer than by helioseismic experiments.  The outcome of these calculations is summarised in Figure 6, For frequencies $< 2\times 10^{-4}$ Hz the $m=2$ (and $m=1$) quadrupole modes are more readily detected in by a LISA type interferometer than by helioseismic experiments.
\vskip 15pt
{\bf 2. Solar Oscillations and space-time around the Sun}

The solar oscillations are normally expressed in terms of a surface harmonic
and Fourier time decomposition with any variable, eg the displacement $\delta {\bf r}$ expressed in the form
$${\delta {\bf r}({\bf r},t)\over R_\odot} = \sum_{\ell=0}^\infty~\sum_{m=-\ell}^\ell~
\sum_{n=-\infty}^\infty {\bf\zeta}_{n\ell m}(r)\,S_{\ell m}
(\theta,\phi)\,e^{i\omega t}\eqno(1)$$
where $\omega = \omega_{n\ell m}$ are the cyclic eigenfrequencies of modes corresponding to a particular 
surface harmonic $S_{\ell m}(\theta,\phi)$, $n$ labels 
the order of the mode ($|n|$ is essentially the overtone number, the number of nodes in the radial direction), 
$\zeta_{n\ell m}(r)$ the corresponding dimensionless eigenfunctions and $(r,\theta,\phi)$ 
spherical polar coordinates with origin at the centre of the Sun.
The modes are classified as p (pressure) modes with frequencies 
increasing with increasing $n$, and g (gravity) modes with frequencies 
decreasing with increasing $n$, (for clarity we use negative 
$n$ for the g-modes in equation (1)). If the basic unperturbed state 
is spherically symmetric the frequencies are independent of azimuthal 
order $m$. Rotation lifts this degeneracy giving frequencies $\omega_{n\ell m} \approx \omega_{n\ell 0} + m\bar{\Omega}$ where 
$\bar{\Omega}\sim\Omega_\odot\sim 3\times10^{-6}$ rad/sec is a weighted mean of the solar angular velocity. With a frequency resolution of $\Delta\nu \sim 3\times 10^{-8}$ Hz, as envisaged for LISA (with 1 year's observation time), these individual m-value modes should be resolved provided there is sufficient power and the line widths are 
sufficiently narrow. For frequencies $\nu = \omega/2\pi \sim 10^{-4}$ Hz, which is the region of interest in the present analysis,  $\Omega_\odot/\omega \sim 10^{-3}$ and the eigenfunctions of the modes may be taken to be independent of azimuthal order $m$. The solar rotation axis is inclined at an angle $\sim 7^o$ 
to the ecliptic plane, which introduces an additional (small) modulation which is neglected in the present analysis.

Since oscillations with frequencies $\sim 10^{-4}$ Hz have not yet been detected in the solar oscillations, the line widths are unknown; the measured line widths at higher frequencies decrease with decreasing frequency and crude extrapolation from the measured range suggests $\Delta\nu < 3\times 10^{-8}$ Hz at frequencies $\sim 10^{-4}$ Hz and consequently mode lifetimes $> 1$ year.  We shall assume here that the lines are narrower than the frequency resolution and the modes may effectively be considered as monochromatic. Only the quadrupole modes $\ell = 2$ are considered; the external gravitational potential of a multipole of order 
$\ell$ decreases like $1/r^{\ell+1}$ and modes with $\ell=2$ will dominate at 1 a.u. Further since the oscillation velocities are very small compared with the velocity of light only quadrupole radiation will be significant.

\vskip 10pt
The external Newtonian gravitational potential of the oscillating Sun can then be expressed in the equivalent forms
$$U({\bf r},t) =  -G\,\int_\odot {\rho dV\over r}
 = U_0({\bf r})  
-{G\over 6}\,{\cal D}^{\alpha\beta}\,
\nabla_{\alpha\beta}\left(1\over r\right)
= U_0({\bf r}) - G M_\odot R^2_\odot  \sum_{m=-2}^2~\sum_{n=-\infty}^\infty
{J_{nm}\over r^3} S_{2 m}(\theta, \phi)\eqno(2)$$
where $U_0({\bf r})$ is the time independent potential, $S_{2 m}$ surface harmonics of degree $\ell = 2$, $J_{nm}\propto e^{i\omega t}$ 
the dimensionless time dependent quadrupole moments corresponding to
eigenmodes with cyclical frequencies $\omega = \omega_{nm}$.
$${\cal D}^{\alpha\beta} = \int_\odot \left(3 x^\alpha x^\beta 
- \delta^{\alpha\beta} x^\mu x_\mu\right)\rho(t)\,dV \eqno(3)$$
is the quadrupole moment tensor, $x^\alpha$ a Cartesian
coordinate system with $\theta =0$ the $x^3$ axis and $\phi=0$ the $x^1$ axis, $\delta^{\alpha\beta}$ is the Kronecker delta and $\nabla_{\alpha\beta} = \partial^2/\partial x^\alpha \partial x^\beta$. 

We take the $S_{2m}$ as real surface
harmonics normalised to unity over the sphere 
$$S_{2,0} = \sqrt{5\over 4\pi}\,{3\cos^2\theta-1\over 2},~~
S_{2,\pm 1} = \sqrt{15\over 16\pi}\,\sin 2\theta\,
\matrix{\cos\phi\cr \sin\phi\cr},~~
S_{2,\pm 2} = \sqrt{15\over 16\pi}\,\sin^2\theta\,\matrix{\cos 2\phi\cr \sin 2\phi\cr},\eqno(4)$$
introduce the set of trace-free basis tensors ${\cal I}_m^{\alpha\beta}$ (corresponding to surface harmonics) as 
$$
{\cal I}_0=\left(\matrix{1&0&0\cr 0&1&0\cr 0&0&-2\cr}\right)
{\cal I}_1=\left(\matrix{0&0&1\cr 0&0&0\cr 1&0&0\cr}\right)
{\cal I}_{-1}=\left(\matrix{0&0&0\cr 0&0&1\cr 0&1&0\cr}\right)
{\cal I}_2=\left(\matrix{1&0&0\cr 0&-1&0\cr 0&0&0\cr}\right)
{\cal I}_{-2}=\left(\matrix{0&1&0\cr 1&0&0\cr 0&0&0\cr}\right)\eqno(5)$$
and expand the quadrupole moment tensor as
$${\cal D}^{\alpha\beta} = M_\odot R^2_\odot\sum_{n=-\infty}^\infty\,\sum_{m=-2}^2 C_m\,J_{nm}{\cal I}_{m}^{\alpha\beta},~~~{\rm where~~}
C_{0} = -\sqrt{5\over 4\pi},~~C_{m} = \sqrt{15\over 4\pi}
~~m\ne 0\eqno(6)$$
The relationships between the $S_{2m},{\cal I}_m,$ and $J_{nm}$ are derived in Appendix 1.

The properties of the eigenmodes were computed using the (iwr-smoothed) standard Aarhus Solar Model S1
 of Christensen-Dalsgaard et al (1996), with the dimensionless radial eigenfunction $\zeta_n(R)$ normalised to $1$ at the solar surface $r=R_\odot$. The amplitude of oscillating quadrupole moments $J_n$ are given in Figure 1. (We here suppress the subscript $\ell = 2$ and $m$, since the eigensolutions are independent of $m$.) Table 1 gives a summary of the results obtained for the amplitudes of relevant quantities derived in this and subsequent sections: column 1 gives the radial order $n$ of the mode, column 2 the frequency $\nu$, column 3 the horizontal displacement eigenfunction at the solar surface $\zeta_h$, column 4 the quadrupole moment $J$. The remaining columns give the computed signals for these modes as evaluated in the following sections: column 5,6,7 gives the amplitude of the gravitational signal function $S_m$ for $m=0,\pm 1,\pm 2$, and in columns 8, 9 the amplitude of the velocity signal $V_m$ for $m=0,\pm 2 $.  Note these are calculated with the normalisation $\zeta_n(R_\odot)=1$; $\zeta_h, J, S_m, V_m$ scale linearly with the value of $\zeta(R)$.  We observe here that since the $m=\pm 1$ modes are antisymmetric about the equator plane $\theta = \pi/2$, these modes will have no net velocity so $V_m= 0, m=\pm 1$. These modes do however have a gravitational signal.

The sun is rotating with a period $\sim 27$ days so a frame of reference fixed 
relative to the oscillating sun rotates relative to an inertial frame, producing a modulation of the time dependent gravitational field. The solar rotation axis 
is inclined to the ecliptic plane, and hence to the orbit plane of the gravitational detector, producing a further modulation of the signals from the oscillations and a very low frequency signal ($\sim 4\times 10^{-7}$ Hz) from the static quadrupole moment induced by the rotation. As the inclination is small ($\sim 7^o$) we neglect these effects in the present analysis taking the rotation axis of the Sun perpendicular to the orbit plane of the detector.
\vskip 8pt
Since the monopole gravitational field of the Sun is weak 
($GM_\odot/rc^2 \sim 10^{-8}$ at 1 a.u.), and that of the time dependent 
quadrupole moments and any associated gravitational radiation even weaker, 
the metric of space-time in the neighbourhood of the Earth is adequately 
described by the weak field limit as
$$ds^2 = g_{ik} dx^i\,dx^k  = (\eta_{ik} + h_{ik}) dx^i\,dx^k 
= \left(1+{2U\over c^2}\right) c^2 dt^2 - 
\left(1-{2U\over c^2}\right)dx^\alpha dx^\alpha 
+ h^{GW}_{\alpha\beta} dx^\alpha dx^\beta\eqno(7)$$
where $\eta_{ik} =  diag (1,-1,-1,-1)$ is the Minkowski metric,
$|h_{ik}|<<1$, and indeces are raised and lowered using the Minkowski tensor.
Roman indeces $i,k = 0,1,2,3$ whereas Greek indeces $\alpha,\beta = 1,2,3$. 
The $h_{ik}$ have a contribution from the time dependent quadrupole moments 
$$h^N_{ik} = {2\,\delta U\over c^2} \delta_{ik} =
 -{G\,M_\odot R^2_\odot\over 3 c^2}\,\delta_{ik}\,{\cal D}^{\alpha\beta}
\nabla_{\alpha\beta}\left(1\over r\right)
=  -{G\,M_\odot R^2_\odot\over 3 c^2}\,\delta_{ik}\,
\nabla_{\alpha\beta}\left(1\over r\right)\,
\sum_{n=-\infty}^\infty\,\sum_{m=-2}^2 C_m\,J_{nm}{\cal I}_{m}^{\alpha\beta}
\eqno(8)$$
and a space-like contribution from the gravitational quadrupole 
radiation, which with time dependence $\propto e^{i\omega t}$ is
$$h^{GW}_{\alpha\beta} = 
{2G\over 3 c^4 r} {d^2\over dt^2}\left(\tilde{\cal D}_{\alpha\beta}\right) 
=-{2\omega^2 G\over 3 c^4 r} \tilde{\cal D}_{\alpha\beta}
=-{2\omega^2 G\,M_\odot R^2_\odot\over 3 c^4 r}\,\sum_{n=-\infty}^\infty\,
\sum_{m=-2}^2 C_m\,J_{nm}\delta_{\alpha\gamma}\,\delta_{\beta\epsilon}\,
\tilde{\cal I}_{m}^{\gamma\epsilon}
\eqno(9)$$
where $\tilde{\cal D}_{\alpha\beta}$ is the retarded, transverse, trace-free part of 
$\cal{D}_{\alpha\beta}$ (ie in the plane perpendicular to the radial direction
of propagation of the waves) and $\tilde{\cal I}_m^{\alpha\beta}$ are the 
retarded, transverse, trace-free tensors corresponding to the basis tensors 
${\cal I}_m^{\alpha\beta}$.

In this weak field limit the $h_{ik}$ can be considered as a superposition of 
independent components and the effect of each component considered 
independently. In the following section we drop the indeces $n$ which
label the radial order of modes and the index $m$ which signifies
the basis tensor (or equivalently the spherical harmonic).  
\vskip 15pt
{\bf 3. Phase shift in a perturbed space-time}

Consider an electromagnetic wave of frequency $\omega_e$ and wave 4-vector 
$k_i$ propagating along the arm of a detector from an emmitter at 
$x_A^i$ to a receiver at $x_B^i$. The phase $\phi$ of the wave at $x_B^i$ is 
given by
$$\int_A^B k_i dx^i =  (k_{iB}x_B^i - k_{iA}x_A^i) - \int_A^B x^i dk_i
= k_{iA}(x_B^i - x_A^i) + \int_A^B (x_B^i-x^i) dk_i\eqno(10)$$
In Minkowski space the wave vector $k^i$ is constant 
$=k_A^i = \omega_e (1,n^\alpha)/c$ and the phase difference $\phi(B)-\phi(A)$ 
is simply $k_{iA}(x_B^i - x_A^i)$. The second term on the right hand side of equation (1) is identically zero in the Minkowski limit. 

In the time-dependent space time $g_{ik}$ the first order perturbation in phase, due both to the departure from Minkowski space time and the displacement of the receiver B (taking A as fixed) is given by
$$\delta \varphi = k_{iA} \delta(x^i_B-x^i_A) + \int_A^B (x^i_B - x^i) dk_i\eqno(11)$$

We take $A$ as the origin of coordinates so $x_A^i=0$, and take the unperturbed 
ray to be given by $x^i = \lambda k^i$, where $\lambda$ is an affine parameter 
varying from $0$ at $A$, to $\Lambda$ at $B$. Since $k^i = dx^i/d\lambda$, the 
null geodesic equation in this weak field approximation reduces to
$${dk^i\over d\lambda} = 
-\Gamma^i_{mn} k^m k^n \approx {1\over 2} h_{mn,i} k^m k^n\eqno(12)$$
and the last term in equation (11) can be expressed as 
$$\int _0^\Lambda (\Lambda - \lambda) k^i dk_i = {1\over 2}
\int _0^\Lambda (\Lambda - \lambda) h_{mn,i} k^m k^n k^i d\lambda=
{1\over 2}
\int _0^\Lambda (\Lambda - \lambda) {dh_{mn}\over d\lambda} k^m k^n d\lambda\eqno(13)$$

Substituting this result into equation (11), defining $l^i=(x^i_B-x^i_A),~
\delta l^i=\delta(x^i_B-x^i_A)$, and integrating by parts gives
$$\delta \varphi = k_{iA} \delta l^i + \left[{1\over 2} 
(\Lambda-\lambda) h_{mn} k^m k^n\right]_0^\Lambda +
 {1\over 2} \int_0^\Lambda h_{mn} k^m k^n d\lambda
\eqno(14)$$
$\delta l^i$ is given by the geodesic deviation equation 
$${D^2\over D\tau^2}(\delta l^i) =  R^i_{jkm} {dx^j\over d\tau} 
{dx^k\over d\tau} l^k ,~~~{\rm which~reduces~to~}~~
{d^2\over dt^2}(\delta l^i) =  R^i_{oko} l^k\eqno(15)$$
in the slow motion approximation appropriate to the current analysis. Since 
in the weak field approximation
$$R^j_{oko} = \Gamma^j_{ok,o} - \Gamma^j_{oo,k} + \Gamma^j_{oi} \Gamma^i_{ok}
- \Gamma^j_{ki} \Gamma^i_{oo} \approx
{1\over 2}\eta^{ij}\left(h_{ki,oo} - h_{oi,ok} + h_{oo,ik} +h_{ok,io}\right)\eqno(16)$$
it follows that $R^o_{oko} = 0, d^2 \delta l^o/dt^2 = 0$. Recalling that 
$h_{o\alpha}=0,~l^i = l_0^i + \delta l^i$ with $l_0^i = (1,n)$ we obtain
$${d^2\over dt^2}\left(\delta l^\alpha\right)
 = {1\over 2}\left[
{d^2\over dt^2}\left( h_\beta^\alpha\right) - 
c^2 \eta^{\alpha\gamma} \nabla^2_{\gamma \beta} h_{oo}\right] l_0^\beta\eqno(17)$$
where again Greek indeces denote spatial components and $\nabla^2_{\gamma \beta}
=\partial^2/\partial x^\gamma\partial x^\beta$.
Since the time dependence of $h_{ik} \propto e^{i\omega t}$ this equation 
can be integrated to give
$$\delta l^\alpha = {1\over 2} \left( h^\alpha_\beta - {c^2
\eta^{\alpha\gamma}\over \omega^2} \nabla^2_{\gamma \beta} h_{oo}\right)
l_0^\beta\eqno(18)$$

Now since $k^\alpha=\omega_e n^\alpha /c, ~\Lambda=cl/\omega_e$, where 
$\omega_e$ is the frequency of the electromagnetic wave, substitution into 
equation(14) gives
$$\delta \varphi = {\omega_e l\over 2 c} \left( h_{\alpha\beta}  n^\alpha
n^\beta - {c^2
\over \omega^2} n^\alpha n^\beta \nabla^2_{\alpha \beta} h_{oo}\right)
- {\omega_e l\over 2 c} h_{mn} n^m n^n + {\omega_e\over 2}  \int_0^{l/c} h_{mn}
n^m n^n dt\eqno(19)$$
and $l = n_\alpha l_0^\alpha$.

We now write $\delta\varphi = \delta\varphi_N +\delta\varphi_{GW}$ where 
$\delta\varphi_N$ is the contributions from the Newtonian potential $\delta U$ and 
$\delta\varphi_{GW}$ that from gravitational waves and expand the $h_{mn}$ in the
last term by Taylor series to give
$$\delta\varphi_N = - {\omega_e l\over 2 c} h_{oo} - {\omega_e l c\over 2
\omega^2}
n^\alpha n^\beta \nabla^2_{\alpha \beta} h_{oo} 
+{1\over 2}{\omega_e } h_{oo} \int_0^{l/c} e^{i\omega t} dt $$
$$\,~~~~~~~~~~+ {1\over 2}\omega_e n^\alpha \nabla_\alpha h_{oo} 
 \int_0^{l/c} e^{i\omega t} c t dt 
+{1\over 4} \omega_e n^\alpha n^\beta \nabla^2_{\alpha\beta} h_{oo}
 \int_0^{l/c} e^{i\omega t} c^2 t^2 dt + \dots\eqno(20)$$
$$\delta\varphi_{GW} = {1\over 2} \omega_e \int_0^{l/c} h^{GW}_{\alpha\beta}
n^\alpha n^\beta dt \approx {1\over 2} {\omega_e l\over c} n^\alpha n^\beta h^{GW}_{\alpha\beta}\eqno(21)$$
We have here used the fact that in the weak field limit 
$h^N_{\alpha\beta} = h_{oo} \delta_{\alpha\beta}$ (see equation 8) to 
eliminate $h^N_{\alpha\beta}$. Retaining just the leading terms in powers of 
$\omega l/c <<1$ and $l/r <<1$, the phase shift reduces to
$$\delta\varphi = -{\omega_e l\over 2 c} n^\alpha n^\beta \left({c^2\over\omega^2}\nabla^2_{\alpha\beta} h_{oo} - h^{GW}_{\alpha \beta}\right)
={\omega_e l\over c}{G\over 6 \omega^2} n^\alpha n^\beta 
\left({\cal D}^{\gamma\epsilon} \nabla^4_{\alpha\beta\gamma\epsilon} \left(1\over r\right) +
{2\omega^4\over c^4 r} \tilde{\cal D}_{\alpha\beta}\right)
\eqno(22)$$
where $\nabla^4_{\alpha\beta\gamma\delta}
= {\partial^4/\partial x^\alpha \partial x^\beta \partial x^\gamma \partial x^\epsilon}$ and we have used the results (8) and (9) for $h^N_{ik}$ and $h^{GW}_{ik}$. 
Since the unperturbed phase shift $\varphi = \omega_e l/c$ the fractional change in phase shift can be expressed in the form 
$$\left({\delta\varphi\over\varphi}\right) = {G\over 2\omega^2 r^5} \left(T_{\alpha\beta}{\cal D}^{\alpha\beta}  + {2\omega^4 r^4\over 3 c^4} N_{\alpha\beta}\tilde{\cal D}^{\alpha\beta}\right)\eqno(23)$$
where  $N_{\alpha\beta}= n_\alpha n_\beta$, and
$$T_{\alpha\beta} = {1\over 3}\,r^5 N^{\gamma\epsilon}\nabla^4_{\alpha\beta\gamma\epsilon}\left(1\over r\right) =
2 n_\alpha n_\beta - 10\mu (n_\alpha \bar{n}_\beta+\bar{n}_\alpha n_\beta)
+ 5 (7 \mu^2 - 1) \bar{n}_\alpha \bar{n}_\beta\eqno(24)$$
where $\bar{n}^\alpha$ is the unit vector in the radial direction and $\mu = n^{\alpha} \bar{n}_\alpha$. The details of the derivation of $T_{\alpha\beta}$ are given in Appendix 2. 
\vskip 15pt
{\bf 4. Response of the interferometer}

A laser interferometer detector such as LISA consists of 3 arms AB, AC, CB, in circular orbit around the Sun; the interferometer's response is given by the difference in the fractional change in phase shifts of the round trip signals along any two of the arms eg. ABA and ACA, as 
$$s = \left(\delta\varphi\over\varphi\right)_{AB}- \left(\delta\varphi\over\varphi\right)_{AC}=
{G\over 2\omega^2 r^5}
\left( \Delta T_{\alpha\beta}{\cal D}^{\alpha\beta}+ {2\omega^4 r^4\over 3 c^4} \Delta N_{\alpha\beta}\tilde{\cal D}^{\alpha\beta}\right)
\eqno(25)$$
where $\Delta$ indicates the difference between the arms AB and AC. 
 
The response $s$ is a function of the angles $\chi, \iota, \psi, \upsilon$, the quadrupole moments $J_m$ and the frequencu $\omega$. $\chi$ is the angle between the radius vector to the detector and the instantaneous orientation of the reference axes corotating with the sun, $\iota$ the inclinations of the plane of the detector to the orbit plane, $\psi$ the angle between the arm AB and the direction of the orbit and $\upsilon$ the angle between the two arms of the detector.  In the LISA concept $\iota=\upsilon = \pi/3$ and the detector rotates with the same period as it orbits the Sun, $\psi$ decreasing as $\phi$ increases, where $\phi$ is the position angle of the detector on its orbit relative to a fixed inertial frame.

We now express the quadrupole moment tensor in terms of the trace-free basis tensors ${\cal I}_m^{\alpha\beta}$ (see equation 6) and determine the response $S_m$ for each mode of azimuthal order $m$ and frequency $\nu = \omega/2\pi$
$$S_m(\nu,\phi) = {1\over 2}\,C_m\,J_m\,\left(R_\odot\over r\right)^5 \left(\nu_\odot\over\nu\right)^2 \,\left[f^N_m + {2\over 3}\left(\nu\over\nu_r\right)^4\,f^{GW}_m\right]$$
where
$$f^N_m= {\cal I}_m^{\alpha\beta} \Delta T_{\alpha\beta},~
f^{GW}_m=\tilde{\cal I}_m^{\alpha\beta} \Delta N_{\alpha\beta}$$
$$\nu_\odot \equiv {1\over 2\pi}\sqrt{G\,M_\odot\over R^3_\odot} \approx 10^{-4} {\rm Hz},~~~\nu_r \equiv \left(c\over 2\pi r\right) \approx 3\times 10^{-4}{\rm Hz} \eqno(26)$$
and $\tilde{\cal I}_m^{\alpha\beta}$ are the transverse-trace-free components of ${\cal I}_m^{\alpha\beta}$. We note that provided $f_m^N$ and $f_m^{GW}$ are of comparable order the gravitational wave contribution dominates for $\nu>\nu_r\sim 3\times 10^{-4}$Hz.

To determine the functions $f^N_m, f^{GW}_m$ consider the detector to be at the point P on its orbit with orbit angle $\phi$ relative to a fixed inertial frame, and let $\chi$ be the angle between the radius vector to P and the $x^1$ axis of the corotating system $x^\alpha$. We then define local transverse Cartesian coordinates $\xi^\gamma$ at P such that $\xi^1$ is in the outward radial direction, $\xi^2$ in the direction of the orbit and $\xi^3$ perpendicular to the orbit plane. The basis tensors in the $\xi^\gamma$ coordinates are given by
$${\sl I}_m^{\gamma\epsilon} = e^\gamma_\alpha e^\epsilon_\beta {\cal I}_m^{\alpha\beta},~~ {\rm where~~}e^1 = (\cos\chi, \sin\chi, 0),~
e^2 = (-\sin\chi, \cos\chi, 0),~e^3 = (0, 0, 1)\eqno(27)$$
are the unit vectors along the $\xi^\gamma$ axes in the $x^\alpha$ coordinate system. The transverse trace-free tensors $\tilde {\sl I}^{\gamma\epsilon}$ are then given by (see Appendix 3)
$$\tilde {\sl I}_m^{23} = \tilde {\sl I}_m^{32} = {\sl I}_m^{23},~~\tilde {\sl I}_m^{22} = -\tilde {\sl I}_m^{33} = 
{1\over 2}\,({\sl I}_m^{22}-{\sl I}_m^{33}),~~~
\tilde {\sl I}_m^{1 \gamma} = 0\eqno(28)$$ 
The detector lies in a plane inclined at an angle $\iota$ to the orbit plane, the arm AB making an angle $\psi$ to the orbital direction $\xi^2$, and AC an angle $\psi+\upsilon$. In the $\xi^\gamma$ coordinates the unit vectors $n_B^\gamma, n_C^\gamma$ are then:
$$n_B^\gamma = (-\sin\psi\cos\iota,
\cos\psi,\sin\psi\sin\iota),~~n_C^\gamma = (-\sin\psi'\cos\iota,
\cos\psi',\sin\psi'\sin\iota),~\psi'=\psi+\upsilon\eqno(29)$$
The unit radial vector $\bar{n}^\gamma = (1,0,0)$ and after some laborious algebra we determine the tensors $\Delta N_{\gamma\epsilon}, \Delta T_{\gamma\epsilon}$ in the $\xi^\gamma$ coordinates, and hence the functions 
$f^N_m, f^{GW}_m$ defined in equation (26). The solutions for arbitrary $\iota,\upsilon,\psi,\chi$  are given in Appendix 3.

In the LISA concept $\iota=\upsilon = \pi/3$ and the detector rotates with the 
same period as the orbit in the sense that $\psi$ decreases as $\phi$ increases.
so that $\psi = \psi_0 - \phi$ where $\psi_0$ is a constant, the orientation of the detector relative to the orbit plane at the arbitrary zero of
the orbit angle $\phi$. We set $\psi_0$ = 0 and obtain
$$f^{N}_0 ={3\sqrt{3}\over 8}\,\sin(2\psi+\pi/3),~~f^{GW}_0 = {21\sqrt{3}\over 16}\,\sin(2\psi+\pi/3)$$
$$f^N_1 = -3\,\left[\,2\,\cos\chi\,\sin(2\psi+\pi/3)
-\sin\chi\,\cos(2\psi+\pi/3)\,\right],~~f^{GW}_1 = {3\over 2}\,\sin\chi\,\cos(2\psi+\pi/3)$$
$$f^N_2 = -{\sqrt{3}\over 2}\,\left[{25\over 4}\,\cos{2\chi}\,
\sin(2\psi+\pi/3) - 8\,\sin{2\chi}\,\cos(2\psi+\pi/3)\,\right]$$
$$f^{GW}_2 = -{7\sqrt{3}\over 16}\,\cos{2\chi}\,\sin(2\psi+\pi/3)
\eqno(30)$$
The values for $m=-1,-2$ are obtained from those for $m=1,2$ by replacing $\chi$ by $\chi-\pi/2m$.

Relative to a fixed inertial frane at time t the detector is at orbit angle $\phi = \Omega_d t$ and the corotating $x^1$ axis is at an angle $\Omega_\odot t$, where $\Omega_d$ is the angular velocity of the detector around the sun and $\Omega_\odot$ the angular velocity of the Sun relative to this inertial frame.  The angle between the radius vector from the sun to the detector and the rotating $x^1$ axis is therefore $\chi =\Omega_s t$ where $\Omega_s= 2\pi/P_s$
where $P_s\sim 26.75$ days is the synodic period of solar rotation, that is the period relative to a reference frame orbiting the sun at 1 a.u. Substituting into the results (30) gives $f^N_m(\phi)$ and $f^{GW}_m(\phi)$ as a function of the position of the detector on its orbit determined by the orbit phase angle $\phi$. The $f^N_m(\phi)$ are shown in Figures 2a-2c and the $f^{GW}_m(\phi)$ in figures 3a-3c. Since the $m=0$ mode is axially symmetric 
the $f_0(\phi)$ are independent of the rotation of the Sun, the other modes display the modulation of the signal due to solar rotation. Combining the 
contributions from $f^N$ and $f^{GW}$ gives 
$$s_0(\nu,\phi) = {3\sqrt{3}\over 16}\,C_0\,J_0\,{R^5_\odot\over r^5}\,{\nu^2_\odot\over\nu^2}\,\left(1+{7\over 3}\,{\nu^4\over\nu_r^4}\right)\,\sin(2\psi+\pi/3)$$
$$s_1(\nu,\phi) = {3\over 2}\,C_1\,J_1\,{R^5_\odot\over r^5}\,{\nu^2_\odot\over\nu^2}\,\left[\,
\left(1 + {1\over 3}\,{\nu^4\over\nu_r^4}\right) \sin\chi\,\cos(2\psi+\pi/3)
-2\,\cos\chi\,\sin(2\psi+\pi/3)\right]$$
$$s_2(\nu,\phi) = 2\sqrt{3} C_2\,J_2\,{R^5_\odot\over r^5}\,{\nu^2_\odot\over\nu^2}\,
\left[\sin{2\chi}\,\cos(2\psi+\pi/3)
 -{25\over 32}\,\left(1 + {7\over 75}\,{\nu^4\over\nu^4_r}\right)\cos{2\chi}\,\sin(2\psi+\pi/3)\right]\eqno(31)$$
$${\rm where~}~~\psi = -\phi,~~\chi = \phi P_d/P_\odot \approx 13.65\,\phi,~~C_{0} = -\sqrt{5\over 4\pi},~~C_{m} = \sqrt{15\over 4\pi},
~~~m=\pm1,\pm2\eqno(32)$$
Again values for $m=-1,-2$ are obtained from those for $m=1,2$ by replacing $\chi$ by $\chi-\pi/2m$. The results (31) give explicitly the contributions from the Newtonian potential and gravitational waves and, as anticipated above, the gravitational wave contribution dominates when $\nu>\nu_r \sim 3\times 10^{-4}$Hz.  

The amplitude of the signal of given $(m,\nu)$ varies over the course of a solar month and during the year. For an observation time $T \sim 1$ year the 
amplitude is given by the root mean square averaged over $T$. The resulting amplitudes are given in Table 1 (columns 5,6,7) for values of the quadrupole moments given in column 3 corresponding to unit surface radial eigenfunction. These signals scale linearly with the value of the quadrupole moments.
\vskip 15pt
{\bf 5. Surface velocity amplitudes on the Sun}

The solar oscillations of low degree are obtained by taking the power spectrum 
of a time series of measurements of the Doppler shift (or velocity) of a 
given spectral line in the integrated light from the sun - the frequencies are 
the peaks in this power spectrum.  The velocity amplitude of a given quadrupole oscillation is therefore given by integrating the component of the surface velocity in the direction of the observer over the visible solar disc, 
$$V = {{\int k^\alpha v_\alpha\,f(\mu)\,\mu\,dS}\over
{\int f(\mu)\,\mu\,dS}}\eqno(33)$$
where in this section we define $k^\alpha$ as the unit vector in the direction of the observer, $\mu = k^\alpha \hat r_\alpha$ as the cosine of the angle between the unit radius vector $\hat r_\alpha$ and the direction to the observer, and $f(\mu)$ the appropriate solar limb darkening function which incorporates the angular dependence of the intensity of radiation at the solar surface (see for example Allen 1973).  We here take $f(\mu) = a + b\mu$ with $a=0.55, b=0.45$,
which is a reasonable approximation for the whole disc velocity measurements by the GOLF (Turk-Chieze 1998) and BiSON (Chaplin et al 1998) experiments

In a system of spherical coordinates $(r,\,\theta,\,\phi')$ corotating with the sun, the velocity on the solar surface due to an oscillation mode is
$$v_\alpha = (v_r, v_\theta, v_\phi) = \omega\,R_\odot\left(\zeta_R,\,\zeta_h {\partial\over\partial\theta},\,
\zeta_h {1\over\sin\theta} {\partial\over\partial\phi'}\right)\,
S_{2 m}(\theta, \phi')\eqno(34)$$
where $\zeta_R, \zeta_h$ are the dimensionless radial and horizontal displacement eigenfunction at the solar surface $R=R_\odot$.
Let the corotating axes be at an angle $\phi_0$ to the direction to the 
observer and take $\phi = \phi'-\phi_0$ then 
$$k^\alpha v_\alpha = v_r \sin\theta\cos\phi + v_\theta \cos\theta\cos\phi - v_\phi \sin\phi,~~~~\mu = \sin\theta\cos\phi\eqno(35)$$
and evaluating the integral (33) gives the velocity amplitudes
$$V_0 = -\omega\,R_\odot\,\sqrt{5\over 4\pi}\left({(16a+15b)\zeta_r + 6(8a+5b)\zeta_h\over 40(3a+2b)}\right),~~
V_{\pm 2} = \sqrt{3}\,V_0\,\matrix{\cos 2\phi_0\cr \sin 2\phi_0\cr}~~(m=\pm 2)\eqno(37)$$
For $m = \pm 1, S_{2m} \propto \cos\theta$ is antisymmetric about $\theta = \pi/2$ and the integrals (32) for $V_{\pm 1}$ are identically zero. The $m=\pm 1$ modes are not detectable in integrated velocity in the equatorial plane.

The surface horizontal displacement $\zeta_h$ is known in terms of the surface value of the radial displacement eigenfunction $\zeta_R$ from integration of the eigenvalue equations for the oscillation (column 3 of Table 1). Since the normalised value of the quadrupole moment (column 4) is $<<1$ for all modes, it follows from the equations governing the oscillations that $\zeta_h \approx \zeta_R\,g/\omega^2 \,R_\odot$ (see Unno et al 1989).  The velocity amplitudes with $\zeta_R = 1$ are listed in column 3 of Table 1; they scale linearly with the value of $\zeta_R$.
\vskip 15pt
{\bf 6. Background Noise the Gravitational and Helioseismology Experiments}

The expected sensitivity of the LISA experiment for a single 2-arm detector 
has recently been re-evaluated; at low frequencies the dominant contribution is from the acceleration noise $\propto 1/\omega^2$, at intermediate frequencies from shot noise, and at high frequencies the sensitivity declines when the path length along the detector becomes comparable to and greater than the wavelength of the gravitational wave (ESA 1998). In the region of interest this 
background noise level is given by 
$$B_i(\nu) = \left(1.6\times 10^{-41} + 9.2\times 10^{-52}{1\over\nu^4}\right)^{1/2}  /{\sqrt{Hz}}\eqno(38$$
This background noise is normally presented in terms of limits
on a signal that could be detected with one years data at a sampling rate of 
1 sec, with a signal to noise of 5, for an unknown location of the source, and including the inclination between the two arms of the detector. To reproduce 
the limits given in ESA (1998) we divide the result (38) by $\sqrt{T}$ (in secs),
mutiply by $5$ for the S/N, by $\sqrt{5}$ for the unknown direction and divide by $\sin\pi/3$ for the angle between the two arms. This is shown in Figure 4. 

In addition to this instrumental noise there is also a contribution from galactic and extragalactic binary systems, the magnitude of this "confusion noise", $B_b(\nu)$, is uncertain but has been estimated by Bender and Hils (1997); their estimate is given in column 11 of Table 1 and shown in Figure 4.
\vskip 10pt
The background noise in velocity experiments is dominantly from velocity fields on the solar surface (active regions, granulation, meso-granulation, supergranulation) and the cumulative effect of these motions has been estimated by Harvey (1995).  However in contrast to the gravitational case this background
noise has been measured in a number of helioseismology experiments and in Figure 5 we show the background velocity noise determined by the GOLF experiment on SOHO (Gabriel 1997,Turck-Chieze 1998). This is in reasonable agreement with the value obtained by the ground based whole disc networks BiSON (Elseworth et al 1994) - who obtained $900$ cm/sec/$\sqrt{Hz}$ at $\nu=5\times 10^{-4}$ Hz and IRIS (Fossat et al 1996) who obtained  $2\times 10^3$ cm/sec/$\sqrt{Hz}$ at $\nu=10^{-4}$ Hz.  These values are a factor $\sim 5$ below the model predictions of Harvey.  
In the region of interest this backkground noise can be approximated by
$$B_v(\nu) \approx 2\times 10^3\,\left(10^{-4}\over \nu\right)^{1/2}\,cm/sec/\sqrt{Hz} \eqno(39)$$
\vskip 15pt
{\bf 7.  Comparison of detectibility in gravitational and velocity experiments}

We now make the following comparison. We assume a frequency resolution in the power series analysis of $3\times 10^{-8}$ Hz (corresponding to 1 year's observation time) and the value of a velocity signal
from an oscillation mode that has S/N=3 using the above background noise
estimate (39).  This gives a value of the dimensionless radial eigenfunction $\zeta_R$ through equations (37) for the velocity amplitude for each frequency and $m=0,2$. This then determines the amplitude of the quadrupole moment $J$ 
and the gravitational signal strength $S$ obtained by simply scaling $S_m$ in Table 1 by the value of $\zeta_R$. We then compare this gravitational signal strength with the background noise for the LISA experiment computed from (38) under the same assumptions: $\Delta\nu = 3\times 10^{-8}$ Hz, S/N=3. The results are shown in Figure 6.  We see that for frequencies $\nu < 2\times 10^{-4}$ Hz the signal from the m=2 modes are stronger in the gravitational experiment than in the velocity experiment  On this diagram we also show the 
S/N for the gravitational experiment when binary confusion noise is included.
Note that since the averaged velocity for the $m=\pm 1$ modes is zero there
is no limit on the gravitational signal set by the helioseismology experiments.

This comparison between the 2 experiments is of course independent of the assumed S/N in the velocity experiment, what is being compared is the detectibility of an oscillation mode by the two techniques and is essentially the ratio of the S/N for the gravitational experiment to S/N for the velocity experiment.  

We can reverse the comparison and ask what would be the velocity amplitude of modes that are at the margin of detectibility with the gravitational detector, and how does this compare with the solar background noise?
This is done in Figure 7, here we see that the velocity amplitude decreases at low frequencies and for the $m=2$ modes is well below the noise level for a frequency resolution of $30$ nHz. The local minimum at $\sim 4 \times 10^{-4}$Hz
is due to the inclusion of gravitational wave emission.

In the above we have assumed that the signal is monochromatic - or rather that the line is narrower than the frequency resolution.  This is the prediction from extrapolating the p-modes observed by helioseismology at higher frequencies. Since the rotational splitting of the modes is known, signal enhancment techniques can be used,  superposing power in frequency bins separated by $m\Omega_\odot$, as is being done in the search for g-modes in helioseismology by the Phoebus group (Fr\"olich et al 1998).
If the modes have a line width in excess of the frequency resolution the detectibility in both velocity and gravity is correspondingly reduced but the
ratio of S/N for the two experiments remains the same.
\vskip 15pt
{\bf 8. Conclusions}

At low frequencies, $\nu < 2.2\times 10^{-4}$Hz, the quadrupole oscillations of the Sun are more readily detected through their gravitational signals on a LISA type space interferometer than by helioseismology. The emission of gravitational waves by the oscillations contributes to the gravitational signal for frequencies $>3\times 10^{-4}$ Hz.

These low frequency modes are more sensitive to the structure of the solar core than the higher frequency p-modes that have so far been measured by helioseismology experiments. Were these low frequency modes detected they would advance our understanding of the structure of this central core.   The helioseismology experiments measure the surface velocity which is dependent on the detailed structure of the outer layers of the sun, and this has to be subtraced off to give diagnostic information on the solar interior. The interferometer experiments however measure the actual quadrupole moments which are determined by the oscillations in the deep high density solar interior and therefore encode information about the solar interior which is independent of the structure and physics of the outer layers.

The predicted amplitudes of these low $n$ modes is also uncertain. The $p$-modes are thought to be stochastically excited by the convective motions, and if this is the excitation mechanism of the g-modes the predicted amplitudes of the low frequency modes are  very small.  But as was first pointed out by Dilke and Gough (1972) and Christensen-Dalsgaard et al (1974), the steep gradient of $^3$He in the inner half of the sun could also provide an excitation mechanism for these modes which may be damped by parametric resonance with other modes or by mild turbulent diffusion, but still be of sufficient amplitude to give rise to a detectable gravitational signal.

We note also the the frequency of $10^{-4}$ Hz corresponds to a period of the order of $160$ minutes.  There have been repeated suggestions that such a signal
has been seen in helioseismology experiments, many but not all of the claims being withdrawn following more detailed analysis (but see Kotov et al 1997). 
In a recent analysis of the GOLF data Palle et al (1998) place an upper limit
on the velocity amplitude of such a mode $\sim 1$ cm/sec, an estimate 
compatible with searches by the Phoebus group (Fr\"olich et al 1998).
We note that for such a 160 min mode an S/N=1 in the helioseismic velocity experiments corresponds to a S/N=20 in a LISA type experiment.

The search for g-modes in the helioseismology experiments is ongoing and hopefully will detect some modes prior to the launch of LISA.  Were such modes to have been identified then they would provide a valuable known signal on a
space interferometer which could then be used to calibrate and test the experiment, giving more credence to the interpretation of signals from more distant astrophysical sources.

\vskip 15pt
{\bf References}
\ref
Allen C W 1973, Astrophysical Quantities, p 169-171, Atlone Press, London
\ref
Bender P and Hils G, 1997. {\it Classical \& Quantum Gravity}, {\bf 14},1439
\ref
Christensen-Dalsgaard J, Dilke F W W \& Gough D O, 1974. {\it Mon Not Roy astr Soc} {\bf 169}, 429
\ref
Christensen-Dalsgaard J, Dappen W, Antia H M, Ajukov S V, Andersen E R,  Basu S, Baturin V A Berthomieu G,  Chaboyer B, Chitre S M, Cox A N, Demarque P, Dziembowski W A, Gabriel M,  Gough D O, Guenther D B,  Guzik J A, Harvey J W, Hill F, Houdek G, Inglesiass C A, Kosovichev A G,  Leibacher J W, Morel P, Proffit C  R, Provost J, Reiter J, Rhodes Jnr E J, Rogers F J, Rogl J, Roxburgh, I W,  Thompson M J, Ulrich R K. 1996.   The Current State of Solar Modeling.  {\it Science}  {\bf 272}, 1296-1300.
\ref
Chaplin W J, Elseworth Y, Isaak G R, Lines R, McLeod C P, Miller B A \& New R, 1998, {\it Mon Not Roy astr Soc} {\bf 298}, L7-L11
\ref
Claverie A, Isaak G R, McLeod C P, van der Raay \& Rocca Cortes T, 1979.
{\it Nature}, {\bf 282} 591
\ref
Cutler C. and Lindblom L., 1996. {\it Phys. Rev. D.}, {\bf 54}, 1287.
\ref
Dilke F W W and Gough D O, 1972.  {\it Nature}, {\bf 240}, 262.
\ref
Elsworth Y, Howe R, Isaak G, McLeod C, Miller B, New R, Speake C, Wheeler S, 1994, {\it Mon Not Roy astr Soc}, {\bf 269}, 529
\ref
ESA 1998. {\it LISA, Laser Interferometer Space Antenna for
Gravitational Wave Measurements: Pre Phase A Study Report} (ESTEC,
ESA, 1998).
\ref
Fr\"ohlich C, Finisterle W, Andersen B, Appourchaux T, Chaplin W J, Elseworth Y, Gough D O, Hoeksema J T, Isaak G R, Kosovichev A G, Provost J, Scherrer P H, Sekki T \& Toutin T, 1998. In {\it Structure and Dynamics of the Interior of the Sun and Sun-Like Stars}, 67-71. Proc. SOHO 6/GONG 98 Workshop, ESA SP-418.
\ref
Fossat E 1996 (private communication)
\ref
Gabriel A H et al, 1997,
in {\it First results from SOHO}, ed B Leck \& Z Svestka, 207-226, Kluwer Academic Publishers, Dordecht, Netherlands.
\ref
Grec G, Fossat E \& Pommerantz M, 1980. {\it Nature}, {\bf 288}, 541
\ref
Harvey J, 1995. ESA SP-235, p199
\ref
Kotov V A, Haynechuck V I, Tsap T T \& Hoeksema J T, 1997.
{\it Solar Physics} {\bf 176}, 45
\ref
Leighton R B, Noyes R W, Simon G W, 1962. {\it Astrophys J}, {\bf 132} 474
\ref 
Palle P L, Rocca Cortes T, Gelly B, Garcia R A, Gabriel A, Grec G, Robillot J M, Turck-Chieze S \& Ultich R K, 1998. In {\it Structure and Dynamics of the Interior of the Sun and Sun-Like Stars}, 291-294. Proc. SOHO 6/GONG 98 Workshop, ESA SP-418.
\ref
Polnarev A G, Giampieri G,  Marchenkov K I, Roxburgh I W \& Vorontsov S V, 1996. The Gravitational Field of Solar Oscillations and the Impact on Space Experiments to Detect Gravitational Radiation. General Relativity XIV. Israel, 1996
\ref
Roxburgh I W, Giampieri G, Polnarev A G, Vorontsov S V, 1999.  The Effect of Solar oscillations on space gravitational wave experiments.
 In {\it The Non-Sleeping Universe}, eds M. T. V. T. Lago \& A. Blanchard, Astrophysics and Space Science Library, p35, Kluwer, Dordrecht.
\ref
Turck-Chieze S,1998. (Private communication) - see also Gabriel A H 1997,
\ref
Unno W, Osaki Y, Ando A, Shibahashi H, 1989, {\it Nonradial Oscillation of Stars},
University of Tokyo Press. 


\vfill\eject
\vskip 15pt
{\bf Appendix 1. The quadrupole basis tensors ${\cal I}^{\alpha\beta}$ and surface harmonics $S_{2 m}$}

The quadrupole gravitational potential can be expressed in the two equivalent forms
$$G\,M_\odot\,R_\odot^2 \sum_{m=-2}^2~\sum_{n=-\infty}^\infty
{J_{nm}\over r^3} S_{2m}(\theta, \phi) = 
G\,{1\over 6}\,{\cal D}^{\alpha\beta}\, \nabla_{\alpha\beta}\left(1\over r\right)\eqno(A1)$$
where $(r,\theta,\phi)$ are spherical polar coordinates, $x^\alpha = (x,y,z)$ Cartesian coordinates with $\theta =0$ the $z$ axis and $\phi = 0$ the $x$ axis,  and $\nabla_{\alpha\beta}=\partial^2 /\partial x^\alpha\partial x^\beta$.  $J_{nm}$ are the (dimensionless) quadrupole moments, ${\cal D}^{\alpha\beta}$ the quadrupole moment tensor and $S_{2m}$ are the real surface harmonics normalised to unity over the sphere
$$S_{2,0} = \sqrt{5\over 4\pi}\,{3\cos^2\theta-1\over 2},~~
S_{2,\pm 1} = \sqrt{15\over 16\pi}\,\sin 2\theta\,
\matrix{\cos\phi\cr \sin\phi\cr},~~
S_{2,\pm 2} = \sqrt{15\over 16\pi}\,\sin^2\theta\,\matrix{\cos 2\phi\cr \sin 2\phi\cr}\eqno(A2)$$
We define a set of 5 independent trace free tensors ${\cal I}_m^{\alpha\beta}$ as
$${\cal I}_0=\left(\matrix{1&0&0\cr 0&1&0\cr 0&0&-2\cr}\right)
{\cal I}_1=\left(\matrix{0&0&1\cr 0&0&0\cr 1&0&0\cr}\right)
{\cal I}_{-1}=\left(\matrix{0&0&0\cr 0&0&1\cr 0&1&0\cr}\right)
{\cal I}_2=\left(\matrix{1&0&0\cr 0&-1&0\cr 0&0&0\cr}\right)
{\cal I}_{-2}=\left(\matrix{0&1&0\cr 1&0&0\cr 0&0&0\cr}\right)\eqno(A3)$$
Now since
$$\nabla_{\alpha\beta}\left(1\over r\right) ={\partial^2\over\partial x^\alpha\partial x^\beta}
\left(1\over r\right) = \left(3 x_\alpha x_\beta - r^2 \delta_{\alpha\beta}\over r^5 \right)\eqno(A4)$$
where $\delta_{\alpha\beta}$ is the Kronecker delta, and 
$$x^1 = x = r\sin\theta\cos\phi,~~x^2 = y = r\sin\theta\sin\phi,~~x^3 = z = r\cos\theta,\eqno(A5)$$
the ${\cal I}_m^{\alpha\beta}$ satisfy the relations
$${\cal I}_0^{\alpha\beta}\,\nabla_{\alpha\beta}\left(1\over r\right) = {3\over r^5}\,(x^2+y^2-2 z^2)
 = {6\over r^3}\,{1 - 3 \cos^2\theta\over 2} = -{6\over r^3}\,\sqrt{4\pi\over 5}\,S_{2,0}\eqno(A6)$$
$${\cal I}_1^{\alpha\beta}\,\nabla_{\alpha\beta}\left(1\over r\right) = {3\over r^5}\,2xz = {6\over r^3}\,\sin\theta\cos\theta\cos\phi = {3\over r^3}\,\sqrt{16\pi\over 15}\,S_{2,1}\eqno(A7)$$
$${\cal I}_{-1}^{\alpha\beta}\,\nabla_{\alpha\beta}\left(1\over r\right) = {3\over r^5}\,2yz = {6\over r^3}\,\sin\theta\cos\theta\sin\phi = 
{3\over r^3}\,\sqrt{16\pi\over 15}\,S_{2,-1}\eqno(A8)$$
$${\cal I}_2^{\alpha\beta}\,\nabla_{\alpha\beta}\left(1\over r\right) = {3\over r^5}\,(x^2-y^2) =  {3\over r^3}\,\sin^2\theta\cos 2\phi
={3\over r^3}\,\sqrt{16\pi\over 15}\,S_{2,2}\eqno(A9)$$
$${\cal I}_{-2}^{\alpha\beta}\,\nabla_{\alpha\beta}\left(1\over r\right) = {3\over r^5}\,2xy =  {3\over r^3}\,\sin^2\theta\sin 2\phi
={3\over r^3}\,\sqrt{16\pi\over 15}\,S_{2,-2}
\eqno(A10)$$
If we now expand the quadrupole moment tensor as
$${\cal D}^{\alpha\beta} = M_\odot R^2_\odot\sum_{n=-\infty}^\infty\,\sum_{m=-2}^2 C_m\,J_{nm}{\cal I}_{m}^{\alpha\beta}\eqno(A11)$$
then on substituting into eqn (A1) the coefficients $C_m$ are found to be
$$C_{0} = -\sqrt{5\over 4\pi},~~C_{m} = \sqrt{15\over 4\pi}
~~m=\pm1,\pm2\eqno(A12)$$
\vfill\eject
{\bf Appendix 2. Derivation of the detector tensor $T_{\epsilon\gamma}$}

We use a cartesian coordinate system $x^\alpha$ with $x^\alpha x_\alpha = r^2$.
Then
$$\nabla_{\epsilon} \left(1\over r\right) = -{x_\epsilon\over r^3}$$
$$\nabla_{\gamma\epsilon} \left(1\over r\right) = -{\delta_{\epsilon\gamma}\over r^3}
+ 3{ x_\epsilon x_\gamma\over r^5}$$
$$\nabla_{\beta\gamma\epsilon} \left(1\over r\right) =
 +3{\delta_{\epsilon\gamma} x_\beta\over r^5}
+ 3{\delta_{\epsilon\beta} x_\gamma\over r^5}
+ 3{x_\epsilon \delta_{\gamma\beta}\over r^5}
-15 {x_\epsilon x_\gamma x_\beta\over r^7}$$
$$\nabla^4_{\alpha\beta\gamma\epsilon} \left(1\over r\right) =
{3\over r^5}(\delta_{\epsilon\gamma} \delta_{\beta\alpha}
+ \delta_{\epsilon\beta} \delta_{\gamma\alpha}
+ \delta_{\epsilon\alpha} \delta_{\gamma\beta})$$
$$~~~~~~~~~~~~ - {15\over r^7}(\delta_{\epsilon\gamma} x_\beta x_\alpha
+ \delta_{\epsilon\beta} x_\gamma x_\alpha
+ \delta_{\epsilon\alpha} x_\gamma x_\beta
+ \delta_{\gamma\beta} x_\epsilon x_\alpha
+ \delta_{\gamma\alpha} x_\epsilon x_\beta
+ \delta_{\beta\alpha} x_\epsilon x_\gamma)
+ {105\over r^9} x_\epsilon x_\gamma x_\beta x_\alpha$$
Defining the unit vector in the radial direction
$$\bar{n}_\epsilon = ({x_1\over r},\,{x_2\over r},\,{x_3\over r})$$
then
$$\nabla^4_{\alpha\beta\gamma\epsilon} \left(1\over r\right) =
{1\over r^5}{\bigg(}\,3\,(\delta_{\epsilon\gamma} \delta_{\beta\alpha}
+ \delta_{\epsilon\beta} \delta_{\gamma\alpha}
+ \delta_{\epsilon\alpha} \delta_{\gamma\beta})$$
$$~~~~~~~~~~~~ - 15\,(\delta_{\epsilon\gamma}\bar{n}_\beta \bar{n}_\alpha
+ \delta_{\epsilon\beta}\bar{n}_\gamma \bar{n}_\alpha
+ \delta_{\epsilon\alpha}\bar{n}_\gamma \bar{n}_\beta
+ \delta_{\gamma\beta}\bar{n}_\epsilon \bar{n}_\alpha
+ \delta_{\gamma\alpha}\bar{n}_\epsilon \bar{n}_\beta
+ \delta_{\beta\alpha}\bar{n}_\epsilon \bar{n}_\gamma)
 + 105\,\bar{n}_\epsilon \bar{n}_\gamma \bar{n}_\beta \bar{n}_\alpha {\bigg)}$$
Defining $\mu = n^\epsilon n_\epsilon$ and contracting with $n^\epsilon n^\gamma$ gives
$$n^\epsilon n^\gamma \nabla^4_{\alpha\beta\gamma\epsilon} \left(1\over r\right) =
{3\over r^5}{\bigg(}\,\delta_{\alpha\beta}
+ n_{\beta} n_{\alpha}
+ n_{\alpha} n_{\beta}$$
$$~~~~~~~~~~~~ - 5\,(\bar{n}_\beta \bar{n}_\alpha
+ \mu n_\beta \bar{n}_\alpha
+ \mu n_\alpha \bar{n}_\beta
+ \mu n_\beta \bar{n}_\alpha
+ \mu n_\alpha \bar{n}_\beta
+ \mu^2 \delta_{\alpha\beta})
+ 35\,\mu^2 \bar{n}_\beta \bar{n}_\alpha {\bigg)}$$
Since this tensor is to be contracted with the symmetric trace free tensor 
${\cal D}^{\beta\alpha} = {\cal D}^{\alpha\beta}$  and $\delta_{\beta\alpha} {\cal D}^{\beta\alpha} = 0$ we only need the trace fee component of this so the terms with $\delta_{\alpha\beta}$ can be removed
leaving
$$T_{\alpha\beta} = {r^5\over 3}n^\epsilon n^\gamma  \nabla^4_{\alpha\beta\gamma\delta} \left(1\over r\right) = 
2 n_{\beta} n_{\alpha} - 10\mu n_\beta \bar{n}_\alpha
- 10\mu n_\alpha \bar{n}_\beta
+ 5 (7 \mu^2 - 1) \bar{n}_\beta \bar{n}_\alpha$$

\vfil\eject
{\bf Appendix 3. The functions $f_j^{N}(\phi)$, $f_j^{GW}(\phi)$.}

Here we compute the source tensor in local transverse Cartesian coordinates $\xi^\alpha$ at a point $P$ on the orbit of the detector  
where $\xi^1,\xi^2$ are in the orbit plane, $\xi^1$ in the the outward radial direction, $\xi^2$ is in the direction of motion and $\xi^3$ perpendicular to the orbit plane. $\chi$ is the angle between $\xi^1$ and the $x^1$ direction of the coordinate system $x^\alpha$ in which the multipole moments are determined.
The unit vectors along the $\xi^\alpha$ axes in the $x^\alpha$ coordinate system are
$$e_\xi^1 = (\cos\chi, \sin\chi, 0),~~
e_\xi^2 = (-\sin\chi, \cos\chi, 0),~~
e_\xi^3 = (0, 0, 1)$$
Hence the basis quadrupole tensors in the $\xi$ coordinate system are given 
by ${\sl I}_m^{\gamma\epsilon} = e^\gamma_\alpha e^\epsilon_\beta {\cal I}^{\alpha\beta}$ and are
$${\sl I}_0=\left(\matrix{1&0&0\cr 0&1&0\cr 0&0&-2\cr}\right)~
{\sl I}_1=\left(\matrix{0&0&\cos\chi\cr 0&0&-\sin\chi\cr \cos\chi&-\sin\chi&0\cr}\right)~
{\sl I}_{-1}=\left(\matrix{0&0&\sin\chi\cr 0&0&\cos\chi\cr \sin\chi&\cos\chi&0\cr}\right)$$
$$
{\sl I}_2=\left(\matrix{\cos2\chi&-\sin2\chi&0\cr -\sin2\chi&-\cos2\chi&0\cr 0&0&0\cr}\right)~
{\sl I}_{-2}=\left(\matrix{\sin2\chi&\cos2\chi&0\cr \cos2\chi&-\sin2\chi&0\cr 0&0&0\cr}\right)$$

The transverse-trace-free radiation tensors in the $\xi$ coordinates are then
$$\tilde{\sl I}_0={3\over 2}\left(\matrix{0&0&0\cr 0&1&0\cr 0&0&-1\cr}\right)~
\tilde{\sl I}_1=-\sin\chi\left(\matrix{0&0&0\cr 0&0&1\cr 0&1&0\cr}\right)~
\tilde{\sl I}_{-1}=\cos\chi\left(\matrix{0&0&0\cr 0&0&1\cr 0&1&0\cr}\right)$$
$$\tilde{\sl I}_2=-{1\over 2}\cos2\chi\left(\matrix{0&0&0\cr 0&1&0\cr 0&0&-1\cr}\right)~
\tilde{\sl I}_{-2}=-{1\over 2}\sin2\chi\left(\matrix{0&0&0\cr 0&1&0\cr 0&0&-1\cr}\right)$$

Now let the plane of the detector be at an inclination angle $\iota$ to the plane of the orbit ($\iota = \pi/3$ in the LISA experiment), and let the detector arm AB be at an angle $\psi$ to the direction of the orbit ($\xi^2$). In the the $\xi$ coordinate system the the unit vector along the arm of the arm AB of the detector is
$$n^\alpha = (-\sin\psi\cos\iota, \cos\psi, \sin\psi\sin\iota)$$
so the projection tensor $N_{AB}^{\alpha\beta} = n^\alpha n^\beta$ is
$$N_{AB}=\left(\matrix{\sin^2\psi\cos^2\iota&-\sin\psi\cos\psi\cos\iota&
-\sin^2\psi\sin\iota\cos\iota\cr -\sin\psi\cos\psi\cos\iota&\cos^2\psi&\cos\psi\sin\psi\sin\iota\cr -\sin^2\psi\sin\iota\cos\iota&\cos\psi\sin\psi\sin\iota&\sin^2\psi\sin^2\iota\cr}\right)$$
Let the angle between the arms AC and AB of the detector be $\upsilon$ so that AC makes an angle $\psi + \upsilon$ with the $\xi^2$ direction, the projection tensor $N_{AC}^{\alpha\beta}$ is given by the above result with $\psi$ replaced by $\psi + \upsilon$ so the difference
$$\Delta N = N_{AB} - N_{AC} = 
\sin\upsilon\left(\matrix{-\sin(2\psi+\upsilon)\cos^2\iota&\cos(2\psi+\upsilon)\cos\iota&
\sin(2\psi+\upsilon)\sin\iota\cos\iota\cr \cos(2\psi+\upsilon)\cos\iota&\sin(2\psi+\upsilon)&-\cos(2\psi+\upsilon)\sin\iota\cr \sin(2\psi+\upsilon)\sin\iota\cos\iota&-\cos(2\psi+\upsilon)\sin\iota&-\sin(2\psi+\upsilon)\sin^2\iota\cr}\right)$$
The signal functions for gravitational waves $f_j^{GW}$ are given by
$$f^{GW}_m = \Delta N_{\alpha\beta} \tilde{\sl I}_m^{\alpha\beta} =
\sin\upsilon\left(\sin(2\psi+\upsilon)\,\tilde{\sl I}_m^{22} -2\cos(2\psi+\upsilon)\sin\iota\,\tilde{\sl I}_m^{23} - 
\sin(2\psi+\upsilon)\sin^2\iota\,\tilde{\sl I}_m^{33}\right)$$
which gives
$$f^{GW}_0 = ~~1.5\,\sin\upsilon\,(1+\sin^2\iota)\,\sin(2\psi+\upsilon)$$
$$f^{GW}_1 = ~~2\,\sin\upsilon\,\sin\iota\,\sin\chi\,\cos(2\psi+\upsilon)$$
$$f^{GW}_{-1} = -2\,\sin\upsilon\,\sin\iota\,\cos\chi\,\cos(2\psi+\upsilon)$$
$$f^{GW}_{2} = -0.5\,\sin\upsilon\,(1+\sin^2\iota)\,\cos{2\chi}\,\sin(2\psi+\upsilon)$$
$$f^{GW}_{-2} = -0.5\,\sin\upsilon\,(1+\sin^2\iota)\,\sin{2\chi}\,\sin(2\psi+\upsilon))$$
\vfil\eject
The Newtonian signal functions $f^{N}_m=\Delta T_{\alpha\beta} {\sl I}_m^{\alpha\beta}$ where
$$\Delta T_{\alpha\beta} = 2 \Delta N_{\alpha\beta} + 10 \Delta M_{\alpha\beta}
+ 5 \Delta R_{\alpha\beta}$$
with
$$M_{\alpha\beta} = \mu(\bar{n}^\alpha n^\beta + \bar{n}^\beta n^\alpha)
,~~~R_{\alpha\beta} = (7\mu^2  - 1) \bar{n}^\alpha\bar{n}^\beta,~~\mu = \bar{n}^\gamma n_\gamma$$
In the $\xi$ coordinates $\bar{n}^\alpha = (1,0,0)$, and for the detector arm AB $\mu = n^\alpha \bar{n}_\alpha = -\sin\psi \cos\iota$ so
$$M_{AB}=\left(\matrix{2\sin^2\psi\cos^2\iota&-\sin\psi\cos\psi\cos\iota&
-\sin^2\psi\sin\iota\cos\iota\cr -\sin\psi\cos\psi\cos\iota&0&0\cr -\sin^2\psi\sin\iota\cos\iota&0&0\cr}\right)$$
and hence 
$$\Delta M = M_{AB} - M_{AC} =
\sin\upsilon\left(\matrix{-2\sin(2\psi+\upsilon)\cos^2\iota&\cos(2\psi+\upsilon)\cos\iota&
\sin(2\psi+\upsilon)\sin\iota\cos\iota\cr \cos(2\psi+\upsilon)\cos\iota&0&0\cr \sin(2\psi+\upsilon)\sin\iota\cos\iota&0&0\cr}\right)$$
The tensor $R_{AB} = (7\mu^2-1)\bar{n}^\alpha \bar{n}^\beta$ gives
$$\Delta R = -7 \sin\upsilon \sin(2\psi+\upsilon) \cos^2\iota\left(\matrix{1&0&0\cr 0&0&0\cr 0&0&0\cr}\right)$$
So finally the tensor $\Delta T_{\alpha\beta}$ is
$$\Delta T = 
\sin\upsilon\left(\matrix{-17\sin(2\psi+\upsilon)\cos^2\iota&-8 \cos(2\psi+\upsilon)\cos\iota&
-8\sin(2\psi+\upsilon)\sin\iota\cos\iota\cr -8\cos(2\psi+\upsilon)\cos\iota& 2 sin(2\psi+\upsilon)&-2 \cos(2\psi+\upsilon)\sin\iota\cr -8 \sin(2\psi+\upsilon)\sin\iota\cos\iota&-2\cos(2\psi+\upsilon)\sin\iota&-2\sin(2\psi+\upsilon)\sin^2\iota\cr}\right)$$
The signal functions $f^N_m= {\sl I}_m^{\alpha\beta} \Delta T_{\alpha\beta}$ are then
$$f^{N}_0 =\sin\upsilon\,(6-21\cos^2\iota)\,\sin(2\psi+\upsilon)$$
$$f^N_1 = -4\,\sin\upsilon\,\sin\iota\,\left[4\,\cos\iota\,\cos\chi\,\sin(2\psi+\upsilon)
-\sin\chi\,\cos(2\psi+\upsilon)\right]$$
$$f^N_{-1} = -4\,\sin\upsilon\,\sin\iota\,\left[4\,\cos\iota\,\sin\chi\,
\sin(2\psi+\upsilon) + \cos\chi\,\cos(2\psi+\upsilon)\right]$$
$$f^N_2 = -\sin\upsilon\,\left[(17\cos^2\iota +2)\,\cos{2\chi}\,\sin(2\psi+\upsilon) - 
16\,\cos\iota\,\sin{2\chi}\,\cos(2\psi+\upsilon)\right]$$
$$f^N_{-2} = -\sin\upsilon\,\left[(17\cos^2\iota +2)\,\sin{2\chi}\,\sin(2\psi+\upsilon) + 
16\,\cos\iota\,\cos{2\chi}\,\cos(2\psi+\upsilon)\right]$$

setting $\iota=\upsilon=\pi/3$ gives
$$f^{N}_0 ={3\sqrt{3}\over 8}\,\sin(2\psi+\pi/3),~~f^{GW}_0 = {21\sqrt{3}\over 16}\,\sin(2\psi+\pi/3)$$
$$f^N_1 = -3\,\left[\,2\,\cos\chi\,\sin(2\psi+\pi/3)
-\sin\chi\,\cos(2\psi+\pi/3)\,\right],~~f^{GW}_1 = 1.5\,\sin\chi\,\cos(2\psi+\pi/3)$$
$$f^N_2 = -{\sqrt{3}\over 2}\,\left[\,6.25\,\cos{2\chi}\,\sin(2\psi+\pi/3) - 
8\,\sin{2\chi}\,\cos(2\psi+\pi/3)\,\right]$$
$$f^{GW}_2 = -{7\sqrt{3}\over 16}\,\cos{2\chi}\,\sin(2\psi+\pi/3)$$

These can be combined to give
$$f_0 ={3\sqrt{3}\over 8}\,\left(1+{7\omega^4 r^4\over 3 c^4}\right)
 \,\sin(2\psi+\pi/3)$$
$$f_1 =  3\left(1 + {\omega^4 r^4\over 3 c^4}\right) \sin\chi\,\cos(2\psi+\pi/3)
-6\,\cos\chi\,\sin(2\psi+\pi/3)$$
$$f_2 = 4\sqrt{3}\,\sin{2\chi}\,\cos(2\psi+\pi/3)
 -{\sqrt{3}\over 8}\,\left(25 + {7\omega^4 r^4\over 3 c^4}\right)\cos{2\chi}\,\sin(2\psi+\pi/3) $$

\vfill\eject
{\bf Figure captions}

Figure 1\hfill\break
Normalised quadrupole moments $J_{n}$ (in units of $G M R^2$) for quadrupolar solar oscillations with surface displacement $\zeta_n(R)=1$.

Figure 2\hfill\break
Signal function $f^N_m$ for oscillations in the Newtonian potential
(a) $m=0$, (b) $m=\pm 1$, (c) $m=\pm 2$.

Figure 2\hfill\break
Signal function $f^{GW}_m$ for graviational waves produced by solar oscillations
(a) $m=0$, (b) $m=\pm 1$, (c) $m=\pm 2$.

Figure 4\hfill\break
Limits on detectable signal from LISA with signal to noise S/N=5.
(a) Instrumental noise, (b) binary confusion noise (Bender and Hils 1997).

Figure 5\hfill\break
Solar background noise in velocity as determined from the GOLF experiment on SOHO.

Figure 6\hfill\break
Comparison of signal to noise (S/N) in LISA experiment for S/N=3 in solar
velocity experiments, for oscillation modes with $\ell=2, m=2$.

Figure 7\hfill\break
Minimum surface velocity amplitude for signal to be detectable by LISA with S/N=3 for modes with $m=0$ and $m=2$.
\vfill\vfill\eject
\end
\epsfbox[64 50 650 787]{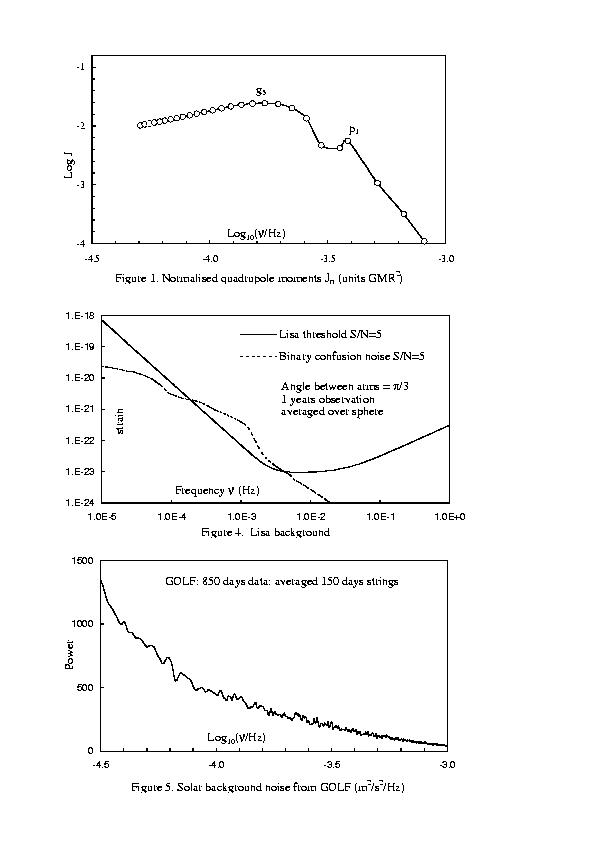}
\vfill\vfill\eject
\epsfbox[85 113 650 776]{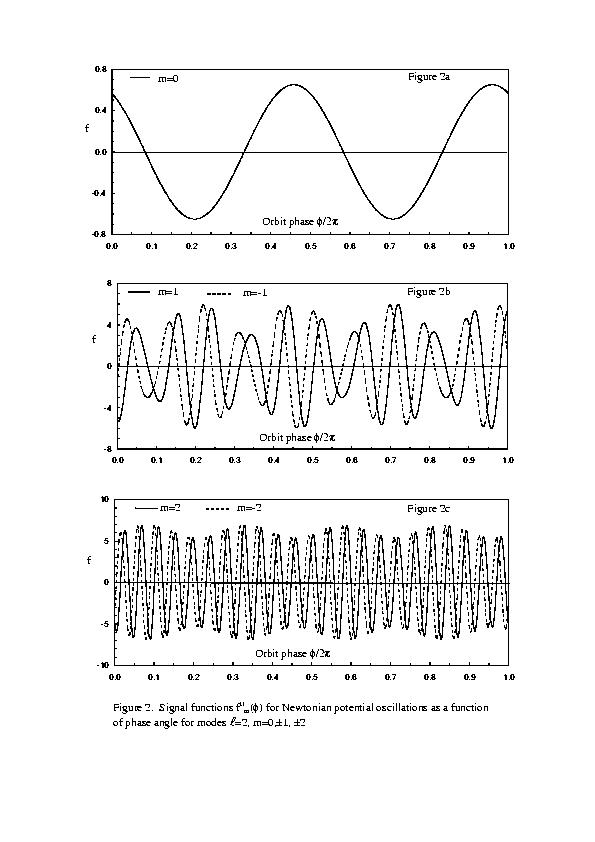}
\vfill\vfill\eject
\epsfbox[88 116 650 776]{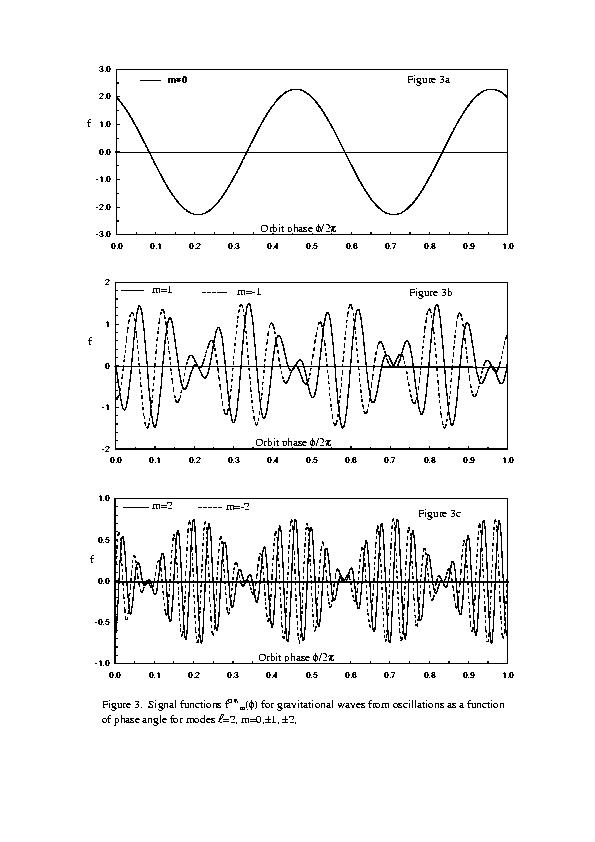}
\vfill\vfill\eject
\epsfbox[70 221 650 780]{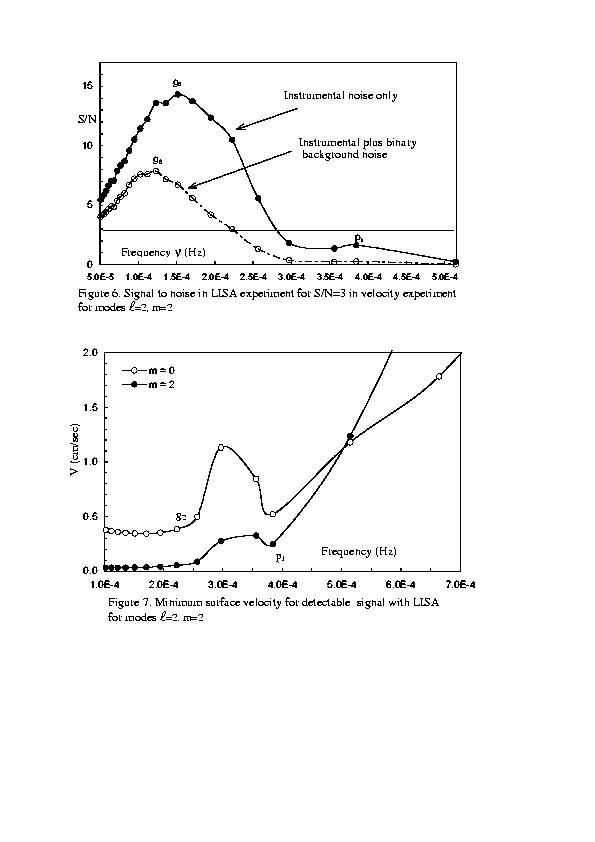}
\vfill\vfill\eject
\epsfbox[109 75 487 768]{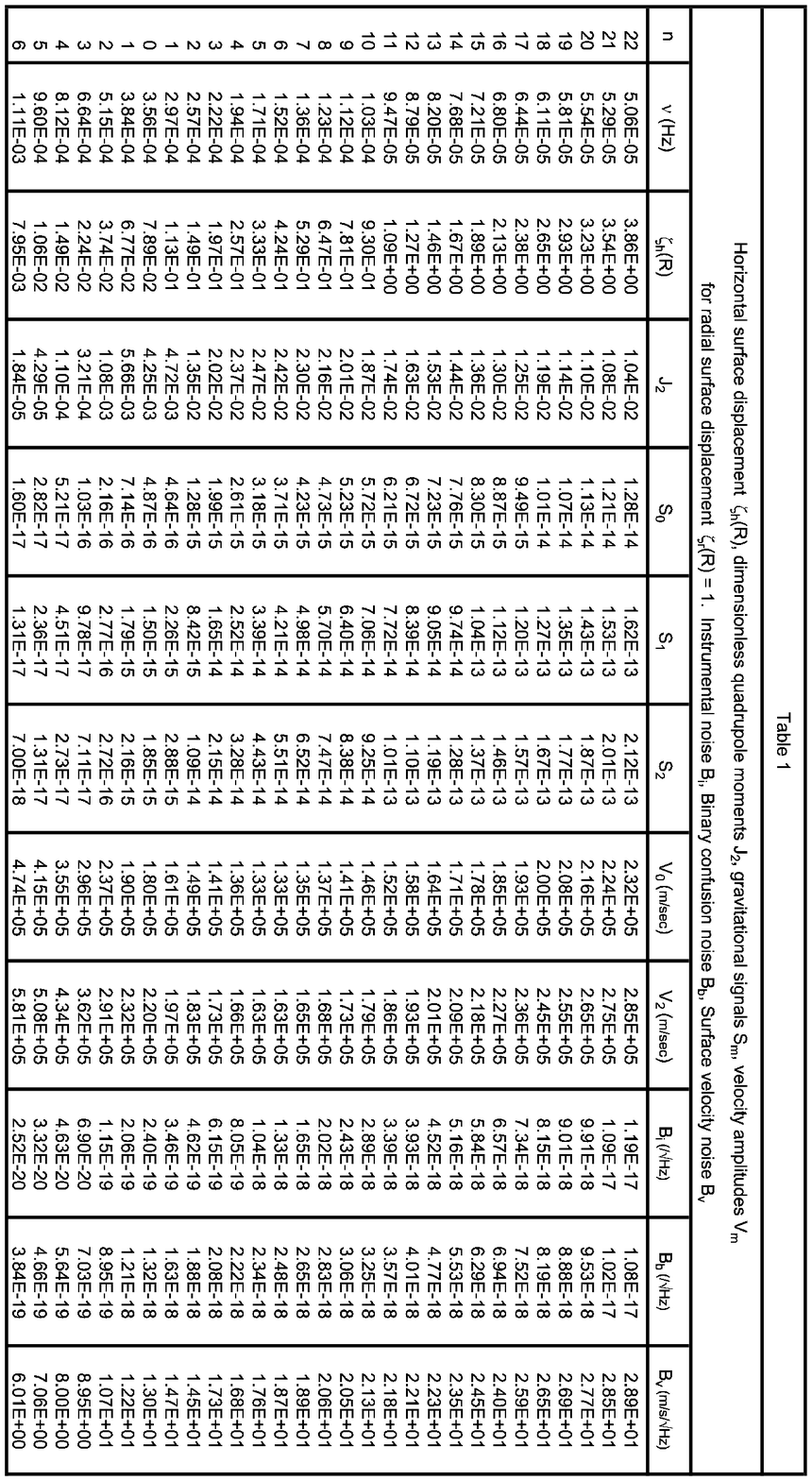}
\vfill\vfill\eject
\end